\documentclass[%
 aip,jcp,
 amsmath,amssymb,
 reprint,
nolongbibliography 
]{revtex4-2}

\usepackage{graphicx}
\usepackage{dcolumn}
\usepackage{bm}

\usepackage[utf8]{inputenc}
\usepackage[T1]{fontenc}
\usepackage{mathptmx}
\usepackage{listings}
\usepackage{rotating} 

\usepackage{physics}

\usepackage{xcolor}
\usepackage{dcolumn} 
\usepackage{bm} 
\usepackage{graphicx}
\usepackage{multirow} 
\usepackage{pifont} 
\usepackage{epsfig}
\usepackage{amsmath} 
\usepackage{subfigure}
\usepackage{float}
\usepackage{booktabs}
\usepackage{tabularx}
\usepackage{natbib}
\usepackage{gensymb}
\usepackage{rotating}
\usepackage{threeparttable}
\usepackage{comment}
\usepackage[normalem]{ulem}
\usepackage[hidelinks]{hyperref} 

\begin{document}
  
\author{Stephen H. Yuwono}
\affiliation{
             Department of Chemistry and Biochemistry,
             Florida State University,
             Tallahassee, FL 32306-4390, USA}
             
\author{A. Eugene DePrince III}
\email{adeprince@fsu.edu}
\affiliation{
             Department of Chemistry and Biochemistry,
             Florida State University,
             Tallahassee, FL 32306-4390, USA}

\title{$N$-Representability Violations in Truncated Equation-of-Motion Coupled-Cluster Methods}

\begin{abstract}

One-electron reduced density matrices (1RDMs) from equation-of-motion (EOM) coupled-cluster with single and double excitations (CCSD) calculations are analyzed to assess their $N$-representability ({\em i.e.}, whether they are derivable from an physical $N$-electron state). We identify EOM-CCSD stationary states whose 1RDMs violate either ensemble-state $N$-representability conditions or pure-state conditions known as generalized Pauli constraints (GPCs). As such, these 1RDMs do not correspond to any physical $N$-electron {\color{black} state}. Unphysical states are also encountered in the course of time-dependent EOM-CC simulations; when an external field drives transitions between a pair of stationary states with pure-state $N$-representable 1RDMs, the 1RDM of the time-dependent state can violate ensemble-state conditions. These observations point to potential challenges in interpreting the results of time-dependent EOM-CCSD simulations.

\end{abstract}

\maketitle

\section{Introduction}

Coupled-cluster (CC) theory\cite{Cizek66_4256,Paldus71_359,Bartlett09_book,Musial07_291} and its excited-state extensions [the closely-related equation-of-motion (EOM)\cite{Bartlett93_7029,Bartlett12_126,Krylov08_433} and linear-response\cite{Monkhorst77_421,Mukherjee79_325,Monkhorst83_1217,Jorgensen90_3333,Helgaker90_3345,Koch97_8059,Jorgensen95_7429,Jorgensen19_134109} CC formalisms] have enjoyed enormous success in quantum chemistry. This success stems from a number of desirable features exhibited by the methods, including the size-extensivity of truncated CC expansions, the size-intensivity of EOM-CC excitation energies, and the rapid and systematic convergence of the {\em ansatz}. 
Despite these nice properties, CC and EOM-CC theories exhibit a few characteristics that could potentially lead to numerical issues. To start, the projected energy associated with a truncated cluster operator ({\em e.g.}, for CC with single and double excitations [CCSD]\cite{Bartlett82_1910}) is not variational; it is possible to obtain a ground-state energy that is lower than the exact (full configuration interaction [CI]) energy within the same one-electron basis set. Second, CC and EOM-CC are not Hermitian theories. As such, a similarity-transformed Hamiltonian ($\bar{H}$) represented in an incomplete many-electron basis can be defective in the sense that its eigenvalues can become complex\cite{Koch17_164105,Gauss21_e1968056} or oscillator strengths computed from its eigenfunctions can be negative,\cite{DePrince23_2305.06412} although these issues can be mitigated by going to a higher-order truncation scheme ({\em e.g.}, CC/EOM-CC with up to triple excitations [CCSDT/EOM-CCSDT]\cite{Bartlett87_7041,Schaefer88_382,Bartlett90_6104,Piecuch01_643,Piecuch01_237,Bartlett01_8263} versus CCSD/EOM-CCSD\cite{Bartlett82_1910,Bartlett93_7029}). In the context of time-dependent (TD) EOM-CC theory,\cite{Head-gordon12_909, Bernhard11_11832, Bernhard11_4678,DePrince16_5834,DePrince17_2951,DePrince19_204107,Li19_6617,Bartlett19_164117,Bartlett21_094103,DePrince21_5438,Koch22_023103,Koch23_2301.05546} such defects manifest as unphysical properties of the time-dependent state, including too-large, negative, or complex-valued or populations of the stationary states that make up the time-dependent superposition.\cite{DePrince23_2305.06412}

These issues reflect a more general problem in truncated EOM-CC methods related to our ability to map eigenfunctions of $\bar{H}$ to a physical $N$-electron state. A specific manifestation of this issue is that, when $\bar{H}$ is expanded in an incomplete many-electron basis, reduced-density matrices (RDMs) associated with its eigenfunctions are not necessarily $N$-representable. In other words, there is no guarantee that the state that {\color{black}produces} such an RDM (or equivalently, a state parametrized by the left- and right-hand EOM-CC wave functions) {\color{black}corresponds to} any physical $N$-electron state. Here, we emphasize that we do not necessarily consider the non-hermiticity of EOM-CC-derived RDMs to be a problem. Rather, we focus on conditions that should be satisfied by the eigenvalues of these RDMs, and we limit our analysis to conditions on the one-particle RDM (1RDM). 

The $N$-representability of the 1RDM can be assessed in two ways. First, ensemble-state $N$-representability, which guarantees that the 1RDM is derivable from an ensemble of $N$-electron density matrices, requires only that the natural spin-orbital occupation numbers (the eigenvalues of the 1RDM) be real-valued and lie between zero and one.\cite{Coleman63_668} Second, pure-state $N$-representability conditions (also called generalized Pauli constraints [GPCs]) guarantee that the 1RDM is derivable from a single $N$-electron density matrix; the GPCs place more complicated,  non-intuitive constraints on the occupation numbers.\cite{Klyachko08_287}

Surprisingly enough, it is easy to identify 1RDMs derived from eigenfunctions of $\bar{H}$ that do not satisfy the basic statistical requirements of ensemble $N$-representability. It is also possible to identify states for which the 1RDM is ensemble $N$-representable but the GPCs are violated. Such states are a bit more rare and tend to appear within the vicinity of complex-valued eigenvalues of $\bar{H}$. What is more interesting is that 1RDMs from non-stationary states encountered in TD-EOM-CC simulations can also violate $N$-representability conditions, and the circumstances under which they do so might not be obvious. For example, consider a ground state from CC and an excited state from EOM-CC that both have pure-state $N$-representable 1RDMs. As we will show below, when an external field drives transitions between these states, the 1RDM for the superposition state can become unphysical, {\color{black}with complex-valued} eigenvalues. 

This manuscript is structured as follows. Section \ref{SEC:THEORY} outlines the general frameworks of CC, EOM-CC, and TD-EOM-CC. The details of our calculations are then provided in Sec.~\ref{SEC:COMPUTATIONAL_DETAILS}. In Sec.~\ref{SEC:RESULTS}, we explore some small systems for which GPCs have been tabulated where 1RDMs from truncated EOM-CC (here, exemplified by EOM-CC with single and double excitations [EOM-CCSD]\cite{Bartlett93_7029}) fail to satisfy the $N$-representability conditions mentioned above. {\color{black}Following some concluding remarks in Sec.~\ref{SEC:CONCLUSIONS}, we present in the Appendix a constrained search formulation of EOM-CC that makes clear that the domain of 1RDMs obtained from solving the truncated EOM-CC eigenvalue problem is larger than the domain of those that satisfy ensemble-state or pure-state $N$-representability conditions.}

\section{Theory}
\label{SEC:THEORY}
 
The ground-state CC wave function is
\begin{equation}
\label{EQN:CC_GS}
    \ket{\Psi_{\rm CC}} = \exp(\hat{T}) \ket{\Phi_0},
\end{equation}
where $\ket{\Phi_0}$ is a Hartree-Fock reference configuration, and $\hat{T}$ is the cluster operator, which, at the CCSD level of theory, takes the form
\begin{equation}
\label{EQN:CLUSTER}
    \hat{T} = \sum_{ia} t_i^a \hat{a}^\dagger_a \hat{a}_i + \frac{1}{4} \sum_{ijab} t_{ij}^{ab} \hat{a}^\dagger_a\hat{a}^\dagger_b \hat{a}_j \hat{a}_i
\end{equation}
Here, $\hat{a}^\dagger$ and $\hat{a}$ are fermionic creation and annihilation operators, respectively, the labels $i, j$ / $a, b$ refer to orthonormal spin orbitals that are occupied / virtual in $\ket{\Phi_0}$, and $t_i^a$ and $t_{ij}^{ab}$ are the cluster amplitudes. The cluster amplitudes are determined by projecting 
\begin{equation}
    \bar{H} \ket{\Phi_0} = E_{\rm CC} \ket{\Phi_0}
\end{equation}
onto the space of configurations that are singly and doubly substituted, relative to $\ket{\Phi_0}$, where $\bar{H} = {\rm exp}(-\hat{T}) \hat{H} {\rm exp}(\hat{T})$ is the similarity-transformed Hamiltonian, and $E_{\rm CC}$ is the ground-state energy.

Given $\hat{T}$, {\color{black}excited states} in EOM-CCSD are defined by the right- and left-hand eigenfunctions of $\bar{H}$:
\begin{align}
    \label{EQN:EOM_CC_eig_R}
    \bar{H} \hat{R}_I \ket{\Phi_0} &= E_I \hat{R}_I \ket{\Phi_0} \\
    \label{EQN:EOM_CC_eig_L}
    \bra{\Phi_0}  \hat{L}_I \bar{H}  &=  \bra{\Phi_0} \hat{L}_I E_I
\end{align} 
where the label {\em I} denotes the state. 
At the EOM-CCSD level, the $\hat{R}_I$ and $\hat{L}_I$ operators are defined by
\begin{equation}
\label{EQN:EOM_CC_R}
    \hat{R}_I ={r}_0 + \sum_{ai} {r}^a_i \hat{a}^\dagger_a\hat{a}_i + \frac{1}{4} \sum_{abij} {r}_{ij}^{ab} \hat{a}^\dagger_a \hat{a}^\dagger_b \hat{a}_j \hat{a}_i 
\end{equation} 
and
\begin{equation}
\label{EQN:EOM_CC_L}
    \hat{L}_I ={l}_0 + \sum_{ai} {l}^i_a \hat{a}^\dagger_i\hat{a}_a + \frac{1}{4} \sum_{abij} {l}_{ab}^{ij} \hat{a}^\dagger_i \hat{a}^\dagger_j \hat{a}_b \hat{a}_a
\end{equation}
respectively, and the right-hand ($r_0$, $r_i^a$, and $r_{ij}^{ab}$) and left-hand ($l_0$, $l_i^a$, and $l_{ij}^{ab}$) amplitudes are determined by solving Eqs.~\ref{EQN:EOM_CC_eig_R} and \ref{EQN:EOM_CC_eig_L}.
Then, the full right-hand and left-hand excited states take the form
\begin{eqnarray}
    \label{EQN:EOM_CC_RIGHT_WFN}
    \ket{\Psi_I} = \hat{R}_I \exp(\hat{T})\ket{\Phi_0} \\
    \label{EQN:EOM_CC_LEFT_WFN}
    \bra{ \tilde{\Psi}_I } = \bra{ \Phi_0} \hat{L}_I \exp(-\hat{T}) 
\end{eqnarray}

At this point, we would like to highlight the asymmetry between the right- and left-hand EOM-CC wave functions. It is well known that the CC wave function (Eq.~\ref{EQN:CC_GS}), even with a truncated cluster operator, spans the same space as the full CI wave function due to the exponential form of the CC {\em ansatz}. {\color{black}The only caveat is that}, for a truncated CC expansion, CI coefficients for Slater determinants with excitation ranks higher than the truncation level of $\hat{T}$ are purely disconnected. Similarly, the right-hand EOM-CC wave functions span the entire Hilbert space, even with truncated $\hat{R}_I$ and $\hat{T}$ operators. On the other hand, the left-hand EOM-CC wave functions do not share this property. The left-hand EOM-CC wave functions are truncated at the same order as $\hat{L}_I$ because the $\exp(-\hat{T})$ part of Eq.~\ref{EQN:EOM_CC_LEFT_WFN} is a de-excitation operator when acting on the bra, which prevents Slater determinants with excitation rank higher than the truncation level of $\hat{L}_I$ from contributing to the overall left-hand EOM-CC wave functions. 
Thus, the wave functions defined by Eqs.~\ref{EQN:EOM_CC_RIGHT_WFN} and \ref{EQN:EOM_CC_LEFT_WFN} do not necessarily span the same space.
This subtlety is not immediately apparent if one views the right- and left-hand EOM-CC wave functions as the right- and left-hand eigenvectors of the {\em truncated} similarity-transformed Hamiltonian as in Eqs.~\ref{EQN:EOM_CC_eig_R} and \ref{EQN:EOM_CC_eig_L}.
Nevertheless, as we show below, this asymmetry in truncated EOM-CC calculations may result in 1RDMs that cannot be derived from a {\color{black}physical} $N$-electron {\color{black}state}.

We also consider a time-dependent formulation of EOM-CCSD in which time-dependent right-hand and left-hand wave functions are given by
\label{EQN:TD_KET}
\begin{equation}
    \ket{\Psi(t)} = \hat{R}(t) \exp(\hat{T}) \ket{\Phi_0}
\end{equation}
and 
\label{EQN:TD_BRA}
\begin{equation}
    \bra{\tilde{\Psi}(t)} = \bra{\Phi_0} \hat{L}(t) \exp(-\hat{T})
\end{equation}
respectively. Here, $\hat{R}(t)$ and $\hat{L}(t)$ have the same structure as the operators given in Eqs.~\ref{EQN:EOM_CC_R} and \ref{EQN:EOM_CC_L}, but the amplitudes are time-dependent. At time $t=0$, these amplitudes are those of the lowest-energy right- and left-hand eigenfunctions of $\bar{H}$, {\em i.e.}, 
\begin{align}
    \hat{R}(0) &= 1 \\
    \hat{L}(0) &= 1 + \sum_{ia} \lambda_a^i \hat{a}_i^{\dagger} \hat{a}_a + \frac{1}{4} \sum_{ijab} \lambda_{ab}^{ij} \hat{a}_i^{\dagger} \hat{a}_j^{\dagger} \hat{a}_b \hat{a}_a    
\end{align}
where the symbols $\lambda^i_a$ and $\lambda^{ij}_{ab}$ represent the usual $\lambda$-amplitudes {\color{black} (or Lagrangian multipliers)} that are obtained by solving the left-hand ground-state CC equations.
Note that amplitudes that make up the cluster operator, $\hat{T}$, are fixed at the values determined for the ground state. The time evolution of the right- and left-hand wave functions are governed by the time-dependent Schr\"{o}dinger equation
\label{EQN:TDEOM_KET}
\begin{equation}
    i \dv{\hat{R}(t)}{t}\ket{\Phi_0} = \left [\bar{H} - \boldsymbol{\bar{\mu}} \cdot \boldsymbol{\epsilon}(t) \right ] \hat{R}(t)\ket{\Phi_0}
\end{equation}
and its complex-conjugate
\label{EQN:TDEOM_BRA}
\begin{equation}
    -i \bra{\Phi_0}\dv{\hat{L}(t)}{t} = \bra{\Phi_0}\hat{L}(t) \left [\bar{H} - \boldsymbol{\bar{\mu}} \cdot \boldsymbol{\epsilon}(t) \right ]
\end{equation}
Here $\boldsymbol{\bar{\mu}} = \exp(-\hat{T}) \boldsymbol{\hat{\mu}} \exp(\hat{T})$ is the similarity-transformed dipole operator, and $\boldsymbol{\epsilon}(t)$ is an applied, time-dependent external electric field.

The elements of the (TD-)EOM-CC 1RDM are defined as
\begin{equation}
    {}^1D^p_q = \bra{\tilde{\Psi}} \hat{a}^\dagger_p \hat{a}_q \ket{\Psi}
\end{equation}
where $\bra{\tilde{\Psi}}$ and $\ket{\Psi}$ represent either left- and right-hand stationary states from EOM-CC or superposition states from TD-EOM-CC, and $p$ and $q$ are general spin-orbital labels. We assess the $N$-representability of the 1RDM by analyzing its eigenvalues, {\color{black}$n_p$}, which correspond to natural spin-orbital occupation numbers. Ensemble $N$-representability conditions are trivial and have a clear physical interpretation; the requirement $0 \le {\color{black} n_p} \le 1$ simply reflects that the probability of a spin-orbital being occupied cannot be less than zero or exceed one (${\color{black}n_p}$ should obviously also be real-valued). The GPCs, on the other hand, place much more stringent restrictions on the occupation numbers. For example, for an ordered set of occupation numbers (${\color{black}n_p}  \ge {\color{black}n_{p+1}}$) corresponding to a system of three electrons distributed among six orbitals (denoted $\wedge^3\mathcal{H}_6$), the GPCs are{\color{black}\cite{Klyachko08_287}}
\begin{align}
{\color{black}n}_1 + {\color{black}n}_6 & = 1\\
{\color{black}n}_2 + {\color{black}n}_5 & = 1\\
{\color{black}n}_3 + {\color{black}n}_4 & = 1\\
{\color{black}n}_4 - {\color{black}n}_5 - {\color{black}n}_6 & \le 0
\end{align}
Clearly, as compared to the ensemble case, it is much more difficult to arrive at physical interpretation of these constraints.  We note that there {\color{black} are combinatorial} GPCs, in general, and they differ for each system $\wedge^N \mathcal{H}_{\color{black}k}$, where ${\color{black}k}$ is the number of spin orbitals. GPCs have only been derived and tabulated\cite{Klyachko08_287} for {\color{black} systems with} up to ten spin orbitals; in the cases we consider below, $\wedge^3 \mathcal{H}_{10}$ and $\wedge^5 \mathcal{H}_{10}$, there are 93 and 161 GPCs, respectively.

\section{Computational Details}
\label{SEC:COMPUTATIONAL_DETAILS}

In Sec.~\ref{SEC:RESULTS}, we analyze 1RDMs derived from EOM-CCSD and TD-EOM-CCSD calculations on a cluster of five hydrogen atoms described by a minimal (STO-3G)\cite{Pople69_2657} basis set. We consider two different charge states, the neutral state (H$_5$) and a doubly-ionized state (H$_5^{2+}$).
We have chosen these systems because these are among the few for which GPCs have been tabulated ($\wedge^5 \mathcal{H}_{10}$ and $\wedge^3 \mathcal{H}_{10}$, respectively).  Spin-orbital CCSD equations and corresponding Python code were generated using the \texttt{p$^\dagger$q} package.\cite{DePrince21_e1954709} Left-hand and right-hand EOM-CCSD wave functions were determined by fully diagonalizing the similarity-transformed Hamiltonian in the basis of the  reference, singly excited, and doubly excited determinants (all with $S_z = +\tfrac{1}{2}$); the relevant equations and Python code were also generated using \texttt{p$^\dagger$q}. TD-EOM-CCSD calculations were carried out in the basis that diagonalizes the similarity-transformed Hamiltonian. All calculations used a restricted open-shell Hartree-Fock (ROHF) reference configuration; ROHF calculations were carried out using \textsc{Psi4},\cite{Sherrill20_184108} and integrals entering the CC equations were taken from \textsc{Psi4}.

\begin{figure}
    \centering
    \includegraphics[width=0.25\textwidth]{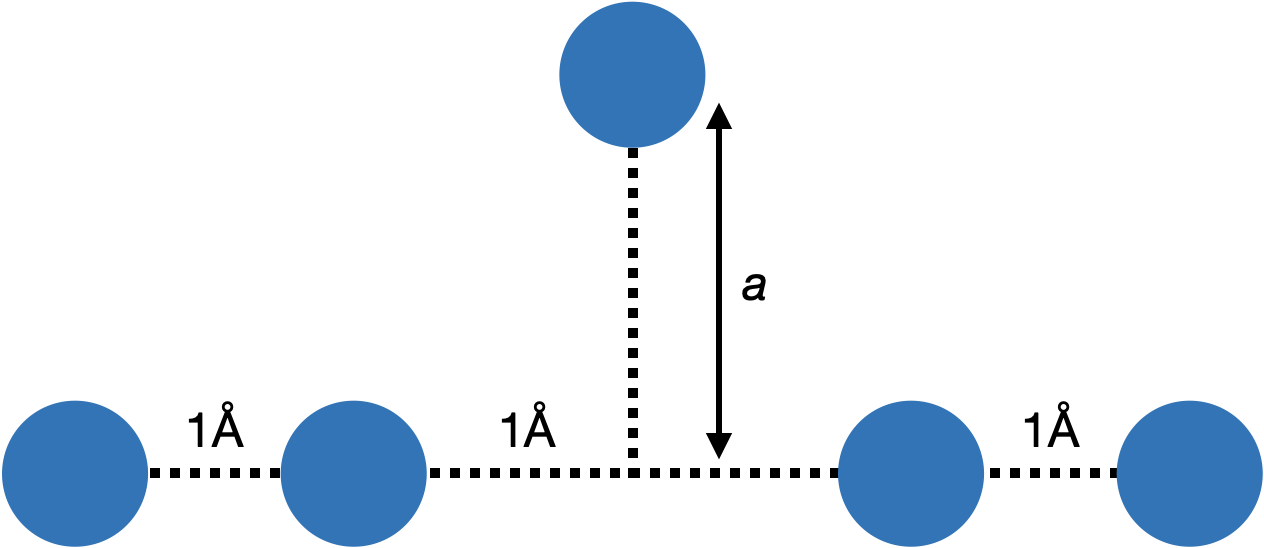}
    \caption{Schematic depicting a linear chain of hydrogen atoms, with an inter-atom separation of 1 \AA. Excited-state potiential energy curves are evaluated along the dissociation pathway defined by moving the central hydrogen atom out of the chain, in a direction perpendicular to the chain. }
    \label{FIG:H5_LINEAR}
\end{figure}

\section{Results}
\label{SEC:RESULTS}

\subsection{H$_5$ / STO-3G}

\begin{figure}
    \centering
    \includegraphics{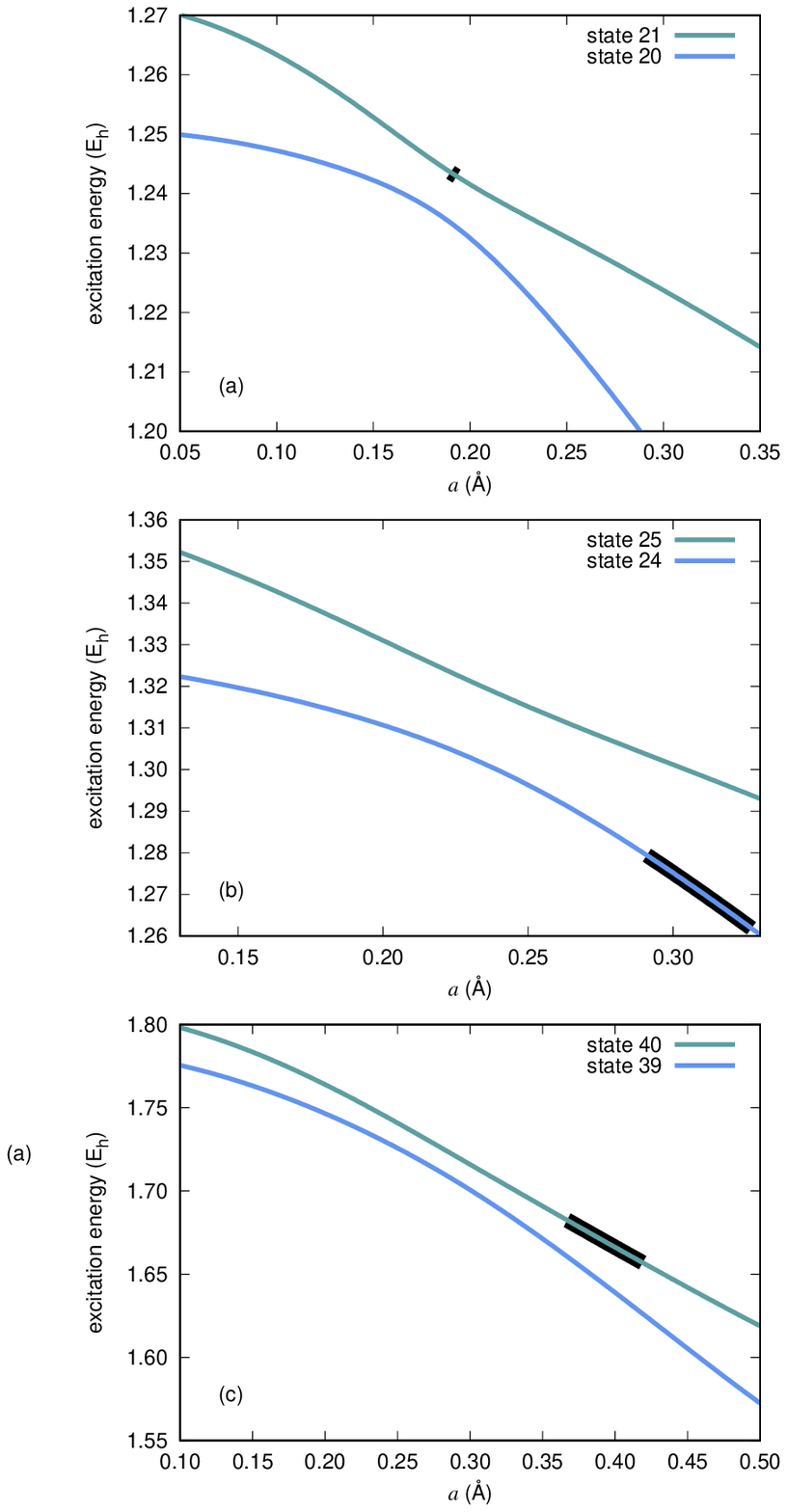}
    \caption{Potential energy curves for the (a) 20$^{th}$ and 21$^{st}$, (b) 24$^{th}$ and 25$^{th}$, and (c) 39$^{th}$ and 40$^{th}$ excited states of of H$_5$.}
    \label{FIG:H5_ONE_STATE_VIOLATES}
\end{figure}

\begin{figure*}
    \centering
    \includegraphics{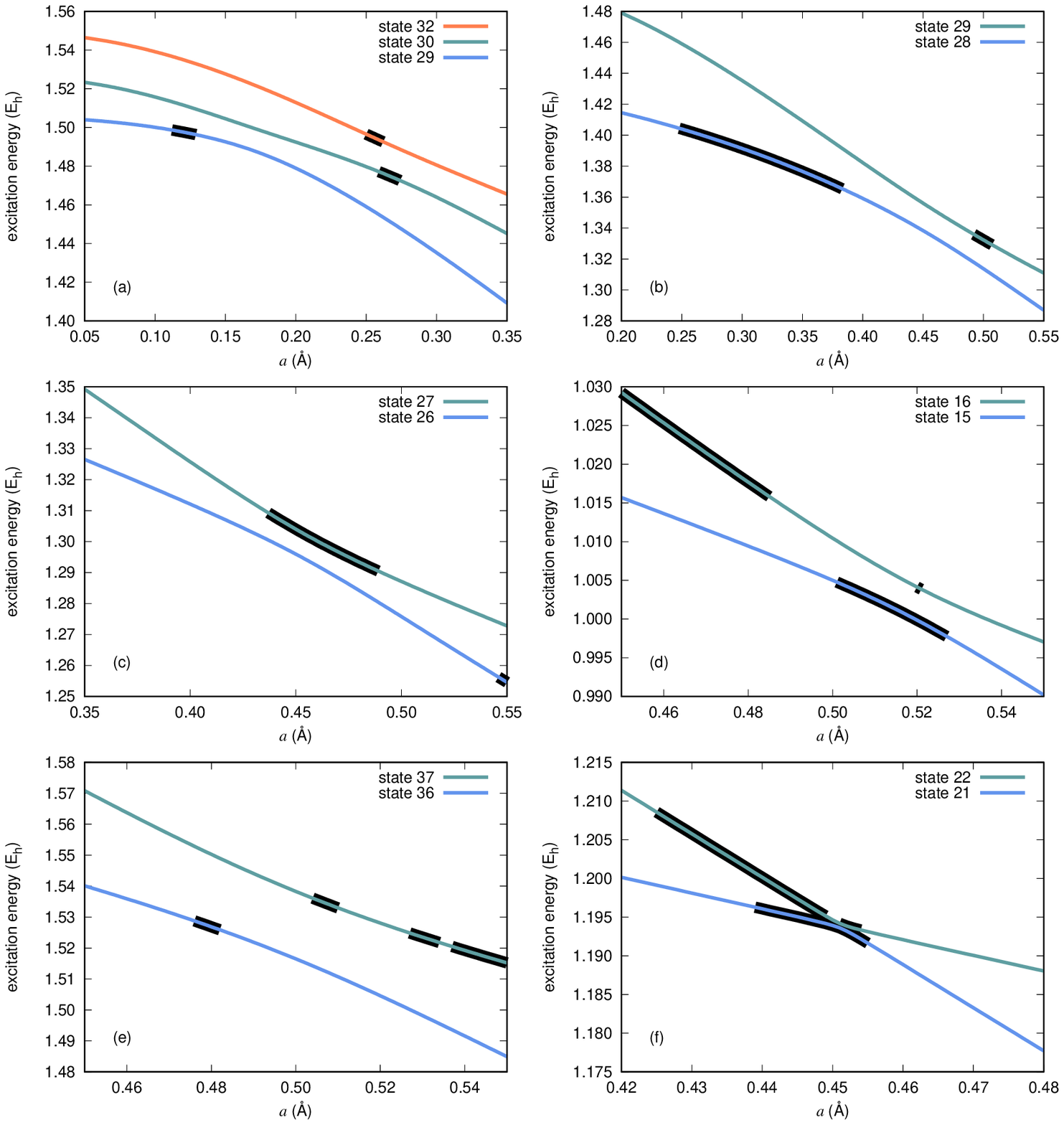}
    \caption{Potential energy curves for the (a) 29$^{th}$, 30$^{th}$, and 32$^{nd}$, (b) 28$^{th}$ and 29$^{th}$, (c) 26$^{th}$ and 27$^{th}$, (d) 15$^{th}$ and 16$^{th}$, (e) 36$^{th}$ and 37$^{th}$, and (f) 21$^{st}$ and 22$^{nd}$ excited states of of H$_5$. State labels refer to the ordering of the states in the depicted windows; the label 29 in panels (a) and (b) refers to the same state.}
    \label{FIG:H5_MULTIPLE_STATES_VIOLATE}
\end{figure*}

\begin{figure}
    \centering
    \includegraphics{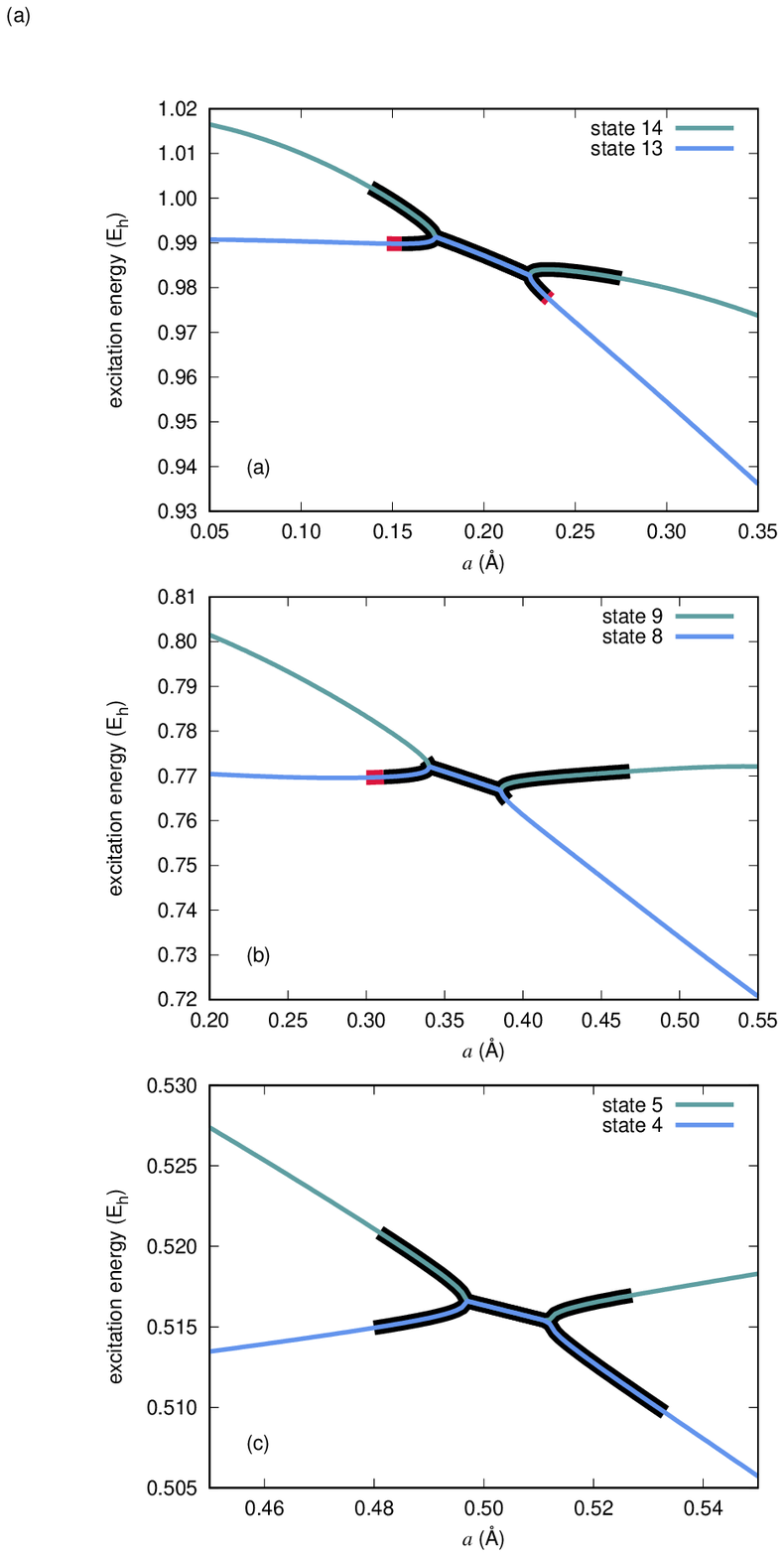}
    \caption{Potential energy curves for the (a) 13$^{th}$ and 14$^{th}$, (b) 8$^{th}$ and 9$^{th}$, and (c) 4$^{th}$ and 5$^{th}$ excited states of of H$_5$. State labels refer to the ordering of the states in the depicted windows.}
    \label{FIG:H5_COMPLEX_ENERGIES}
\end{figure}

We now consider a linear arrangement of five hydrogen atoms (neutral H$_5$), separated by 1 \AA~and described by the STO-3G basis set. We generate potential energy curves along a dissociation path defined by moving the central hydrogen atom out of the chain (by up to 0.55 \AA), in a direction perpendicular to the chain (see Fig.~\ref{FIG:H5_LINEAR}).  Along this pathway, we observe many avoided crossings between excited-state potential energy curves, some of which, based on the energy alone, appear to be well-described by EOM-CCSD (Figs.~\ref{FIG:H5_ONE_STATE_VIOLATES} and \ref{FIG:H5_MULTIPLE_STATES_VIOLATE}), while others exhibit clear problems (Fig.~\ref{FIG:H5_COMPLEX_ENERGIES}).

First, consider the potential energy curves provided in Fig.~\ref{FIG:H5_ONE_STATE_VIOLATES}. Each panel of this figure depicts an avoided crossing between two states, which EOM-CCSD appears to handle correctly,
at least in the sense that the energy remains real-valued. The eigenfunctions of $\bar{H}$, on the other hand, are clearly not physical. Specifically, 1RDMs derived from these eigenfunctions are not $N$-representable in the vicinity of the avoided crossing (depicted by thick black lines), and oscillator strengths sometimes take on negative values.   For example, for state 21, two natural orbital occupation numbers become complex from $a$ = 0.190 \AA~-- 0.193 \AA~(see Fig.~\ref{FIG:H5_LINEAR} for the definition of $a$); the oscillator strength for state 21 also becomes negative (yet small, $\approx -10^{-4}$) beyond $a=0.305$ \AA . Second, two natural orbital occupation numbers for state 24 become complex for $a$ = 0.291 \AA~-- 0.327 \AA; the oscillator strength for this state also becomes negative (yet small, $\approx$ $-10^{-5}$) beyond 0.328 \AA. Third, a pair of natural orbital occupation numbers for state 40 becomes complex for $a$ = 0.367 \AA~-- 0.419 \AA; in this case, though, the oscillator strength for this state remains positive.

Figure \ref{FIG:H5_MULTIPLE_STATES_VIOLATE} provides six additional sets of potential energy curves where EOM-CCSD appears to describe avoided crossings correctly, in terms of the energies, but 1RDMs in the vicinity of the avoided crossings are not physical. In these cases, each state depicted has a non-$N$-representable 1RDM at some point near the avoided crossing. Panel (a) depicts a situation where three states 
interact, and 1RDMs derived from the relevant left- and right-hand eigenfunctions of $\bar{H}$ appear to be defective. Pairs of natural orbital occupation numbers for states 29, 30, and 32 become complex from $a$ = 0.112 \AA~-- 0.129 \AA, $a$ = 0.259 \AA~-- 0.274 \AA, and $a$ = 0.250 \AA~-- 0.262 \AA, respectively. Moreover, oscillator strengths for states 29 and 30 are negative (yet small) for portions of the region under consideration. State 29 has another avoided crossing with state 28 [panel (b)], and both states have pairs of natural orbital occupations that become complex in the vicinity of this avoided crossing. Here, oscillator strengths for state 29 and 28 are slightly negative prior to $a$ = 0.346 \AA~and after $a$ = 0.430 \AA, respectively. The story is similar in panels (c)-(e), where each state exhibits a pair of complex-valued natural orbital occupation numbers, depicted by the black portions of the curves, and states 15, 22, and 37 exhibit slightly negative oscillator strengths in some regions. The most interesting case is the avoided crossing between states 21 and 22 in panel (e). Both states 21 and 22 have a slightly negative oscillator strengths for portions of the depicted region, and both states have pairs of natural orbitals with complex occupation numbers; for portions of the curve, state 22 has four natural orbitals with complex occupation numbers. State 21 also exhibits a real-valued negative occupation number as {\color{black}negative} as {\color{black}$-0.19$} and a real-valued non-negative occupation number as large as 1.04 (both at  $a$ = 0.451 \AA).

We now consider avoided crossings that lead to more obvious deficiencies of EOM-CCSD. Figure \ref{FIG:H5_COMPLEX_ENERGIES} depicts three pairs of EOM-CCSD-derived potential energy curves where the energies become complex (the regions where the real parts are degenerate). In each case, we also observe 1RDMs that do not correspond to any physical $N$-electron state. First, along each complex seam, all natural orbital occupation numbers are complex-valued. Outside of the seam, there exist additional regions where the 1RDMs are still not $N$-representable. In Fig.~\ref{FIG:H5_COMPLEX_ENERGIES}, as above, thick black lines indicate that the 1RDMs for these states do not satisfy ensemble-state $N$-representability conditions. A notable difference here, though, is that we also observe violations in pure-state $N$-representability conditions where the 1RDMs are otherwise ensemble-state $N$-representable [indicated by thick red lines in panels (a) and (b)].

In Fig.~\ref{FIG:H5_COMPLEX_ENERGIES}(a), the 1RDM for state 13 has occupation numbers greater than one for $a$ = 0.157 \AA~-- 0.173 \AA~and negative occupation numbers from $a$ = 0.155 \AA~-- 0.173 \AA. The most extreme values we observe are 3.24 and $-1.74$, just prior to the complex seam at $a$ = 0.173 \AA.  Similarly, the 1RDM for state 14 also has occupation numbers as large as 2.36 or as negative as {\color{black}$-2.20$} prior to the complex seam. After the complex seam ends at $a$ = 0.226 \AA, too large and negative occupation numbers for both states emerge again, and a pair of complex-valued natural orbital occupation numbers appear at some geometries for state 13.  We also observe violations in pure-state $N$-representability conditions for state 13, which are depicted by thick red lines. 
Specifically, we observe violations in the following $\wedge^5 \mathcal{H}_{10}$ GPCs 
\begin{widetext}
\begin{align}
\label{EQN:GPC_108}
3{\color{black}n}_1 +  3 {\color{black}n}_2 + 3 {\color{black}n}_3 +13 {\color{black}n}_4-7{\color{black}n}_5 +  3 {\color{black}n}_6 + 3 {\color{black}n}_7 -7{\color{black}n}_8 -7{\color{black}n}_9 -7 {\color{black}n}_{10} & \le 15 \\
\label{EQN:GPC_99}
7 {\color{black}n}_1  + 7{\color{black}n}_2 +  7{\color{black}n}_3 +   7{\color{black}n}_4 -13 {\color{black}n}_5 -3 {\color{black}n}_6 -3 {\color{black}n}_7 -3 {\color{black}n}_8 -3 {\color{black}n}_9 -3 {\color{black}n}_{10} &\le 15 
\end{align}
\end{widetext}
Equation \ref{EQN:GPC_108} is violated from $a$ = 0.147 \AA~-- 0.154 \AA, and Eq.~\ref{EQN:GPC_99} is violated at $a$ = 0.236 \AA. Note that these violations occur in regions where the 1RDMs are ensemble $N$-representable. Aside from these $N$-representability issues, states 14 and 13 have slightly negative oscillator strengths (on the order of $-10^{-2}$) prior to $a$ = 0.174 \AA~ and from $a$ = 0.226 \AA~-- 0.353 \AA, respectively.

{\color{black}We make similar observations} around the complex seam depicted in Fig.~\ref{FIG:H5_COMPLEX_ENERGIES}(b). The 1RDMs for states 8 and 9 have occupation numbers greater than one or less than zero prior to and after the complex seam. For state 8, the most extreme occupation numbers outside of the seam are 3.48 and $-2.33$; for state 9, we observe occupation numbers as large as 3.16 or as negative as $-2.12$. Along the complex seam ($a$ = 0.341 \AA~-- 0.385 \AA), all natural orbital occupation numbers are complex, and both states occasionally have pairs of complex occupation numbers outside of this region. As with state 13, we also observe pure-state $N$-representability errors for state 8. Specifically, we observe violations in the following $\wedge^5 \mathcal{H}_{10}$ GPCs 
\begin{widetext}
\begin{align}
\label{EQN:GPC_96}
7{\color{black}n}_1 +   7 {\color{black}n}_2 +  7 {\color{black}n}_3 -3 {\color{black}n}_4 -3 {\color{black}n}_5 +  7 {\color{black}n}_6 -13 {\color{black}n}_7 -3 {\color{black}n}_8 -3 {\color{black}n}_9 -3 {\color{black}n}_{10} & \le 15 \\
\label{EQN:GPC_104}
3{\color{black}n}_1 + 13 {\color{black}n}_2 + 3 {\color{black}n}_3 -7{\color{black}n}_4 +  3 {\color{black}n}_5 + 3{\color{black}n}_6  -7 {\color{black}n}_7 + 3{\color{black}n}_8 -7{\color{black}n}_9 -7{\color{black}n}_{10} & \le 15 \\
\label{EQN:GPC_105}
3 {\color{black}n}_1 + 3 {\color{black}n}_2 + 3 {\color{black}n}_3 + 3 {\color{black}n}_4 + 3 {\color{black}n}_5 +13{\color{black}n}_6 -7{\color{black}n}_7 -7{\color{black}n}_8 -7{\color{black}n}_9 -7 {\color{black}n}_{10} & \le 15
\end{align}
\end{widetext}
Equations \ref{EQN:GPC_96}, \ref{EQN:GPC_104}, and \ref{EQN:GPC_105} are violated from $a$ = 0.300 \AA~-- 0.311 \AA, $a$ = 0.302 \AA~-- 0.311 \AA, and $a$ = 0.310 \AA~-- 0.311 \AA, respectively. As before, we stress that these pure-state violations occur in regions where the 1RDMs are ensemble-state $N$-representable. Oscillator strengths for both states are non-negative over the range of $a$ values considered, so these $N$-representablity issues offer the only evidence that the states are unphysical outside of the complex seam.

{\color{black}Figure} \ref{FIG:H5_COMPLEX_ENERGIES}(c) provides potential energy curves for states 4 and 5 in a region where the respective energies become complex ($a$ = 0.497 \AA~-- 0.511 \AA). In this region, all natural orbital occupation numbers are complex, and, as in the other cases, 1RDMs for these states are found to violate ensemble $N$-representability conditions outside of this region. When the energies are real-valued, some occupation numbers are complex valued, while others exceed one (as large as 1.33 and 2.06 for states 4 and 5, respectively) or are less than zero (as negative as {\color{black}$-0.69$} and $-0.83$ for states 4 and 5, respectively). In addition, state 5 has negative oscillator strengths for $a$ < 0.497 \AA~(very negative near the complex root, {\em e.g.}, $< -1.0$, at $a$ = 0.496 \AA). State 4 has negative oscillator strengths for $a$ > 0.511 (again, very negative near the complex root, {\em e.g.}, {\color{black}$< -6.0$} at $a$ = 0.512 \AA). Unlike the other cases where the energy becomes complex, we do not observe any violations in GPCs on the 1RDM for these states at any geometries where the 1RDMs are ensemble-state $N$-representable.

{\color{black}Some of the $N$-representability issues depicted in Figs.~\ref{FIG:H5_ONE_STATE_VIOLATES}--\ref{FIG:H5_COMPLEX_ENERGIES}, specifically the emergence of complex occupation numbers, result from the fact that the 1RDM for a truncated EOM-CC state is not Hermitian. Hence, it stands to reason that some measure of non-Hermiticity of the 1RDM could serve as a proxy for $N$-representability analysis in routine applications of EOM-CC. We have quantified the non-Hermiticity of the 1RDM using the two-norm of the difference between the 1RDM and its adjoint ($||{}^1{\bf D} - {}^1{\bf D}^\dagger||$), and Figs.~S1--S3 in the Supplementary Material depict this quantity for the states considered in Figs.~\ref{FIG:H5_ONE_STATE_VIOLATES}--\ref{FIG:H5_COMPLEX_ENERGIES}. We report several  observations from these data here. First, 1RDMs for each of the excited states considered are significantly more non-Hermitian than the ground-state 1RDMs. Second, the non-Hermiticity spikes in regions having close avoided crossings [see Figs.~\ref{FIG:H5_MULTIPLE_STATES_VIOLATE}(f) and S2(f)] or near complex-valued energy eigenvalues [see Figs.~\ref{FIG:H5_COMPLEX_ENERGIES} and S3]. However, local maxima in non-Hermiticity are not a universal indicator for $N$-representability errors. Several other cases  show that $N$-representability errors persist away from local maxima in the non-Hermiticity [{\em e.g.}, Figs.~S1(c), S2(c), and S2(d)] or when the non-Hermiticity is flat as a function of the geometry [{\em e.g.}, Figs.~S2(a), S2(b), and S2(e)]. Hence, while significant non-Hermiticity can in some cases indicate loss of $N$-representability of the 1RDM, such an analysis should be complemented by one that considers the eigenvalues of the 1RDM themselves to gain a more complete picture of potential problems.
}

\subsection{H$_5^{2+}$ / STO-3G}

\begin{figure}
    \centering
    \includegraphics{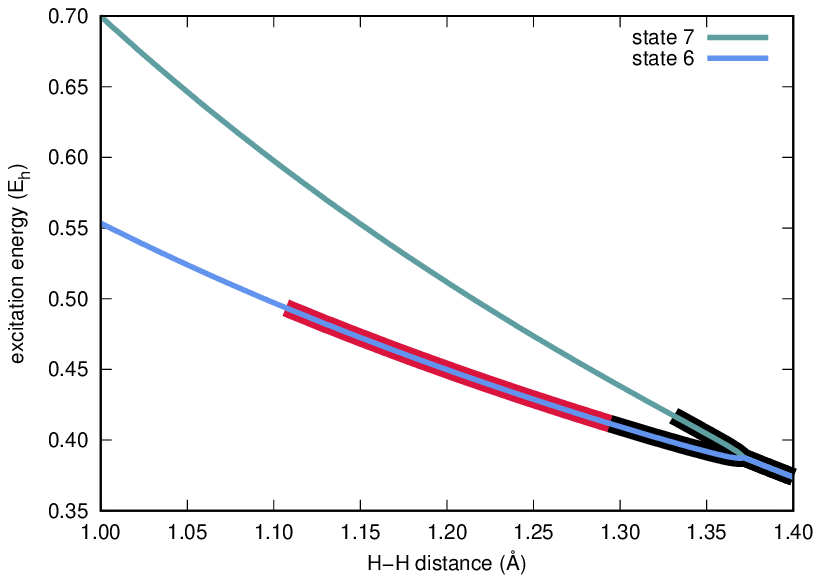}
    \caption{Potential energy curves for the 6$^{th}$ and 7$^{th}$ excited states of H$_5^{2+}$.}
    \label{FIG:H5_2+}
\end{figure}

We now demonstrate that the ensemble-state and pure-state $N$-representability issues discussed above are not limited to systems of the $\wedge^5 \mathcal{H}_{10}$ type. Figure \ref{FIG:H5_2+} shows potential energy curves for the 6$^{th}$ and 7$^{th}$ excited states of H$_5^{2+}$, which falls into the category $\wedge^3 \mathcal{H}_{10}$. Here, the atoms are arranged in a regular pentagon, and the potential energy curves are generated as a function of the distance between adjacent centers. The overall excited-state landscape includes a number of states with non-ensemble-state $N$-representable 1RDMs, but we focus only on one situation in which we identify severe violations in pure-state conditions.

Figure \ref{FIG:H5_2+} demonstrates the onset of a complex seam at an H-H distance of 1.371 \AA, which involves states 6 and 7. As indicated by the thick black lines, the 1RDMs for state 7 violates ensemble-state $N$-representability conditions beyond H-H distances of 1.331 \AA. At this point, one natural orbital occupation exceeds one. At and H-H distance of 1.343 \AA, a pair of occupation numbers becomes complex, and, at a distance of 1.345 \AA, one occupation number becomes negative. Before the complex seam, the most extreme real-valued occupation numbers we observe are 8.5 and -7.8 at an H-H distance of 0.370 \AA. State 6 is more interesting because 1RDMs computed for this state 6 violate several GPCs at shorter H-H distances (1.107 \AA~-- 1.294 \AA), when the 1RDMs are ensemble-state $N$-representable;
the GPCs in question have the following form:
\begin{widetext}
\begin{align}
    \label{EQN:GPC_41}
    9{\color{black}n}_1 +   9{\color{black}n}_2 -11{\color{black}n}_3  -{\color{black}n}_4  -{\color{black}n}_5  -{\color{black}n}_6  -{\color{black}n}_7  -{\color{black}n}_8  -{\color{black}n}_9  -{\color{black}n}_{10} &\le 7 \\
    \label{EQN:GPC_8}
    3{\color{black}n}_1 + 3{\color{black}n}_2 -2{\color{black}n}_3 -2{\color{black}n}_4 + 3{\color{black}n}_5 + 3{\color{black}n}_6 -2{\color{black}n}_7 -2{\color{black}n}_8 -2{\color{black}n}_9 -2{\color{black}n}_{10} &\le 4 \\
    \label{EQN:GPC_30}
    7{\color{black}n}_1 +  7{\color{black}n}_2 -8{\color{black}n}_3 -3{\color{black}n}_4  +2{\color{black}n}_5  +2{\color{black}n}_6 -3{\color{black}n}_7 -3{\color{black}n}_8 -3{\color{black}n}_9 + 2{\color{black}n}_{10} &\le 6 
\end{align}
\end{widetext}
Equation \ref{EQN:GPC_41} is violated from an H-H distance of 1.106 \AA~-- 1.293 \AA. 
At slightly larger H-H distances, two additional GPCs (Eqs.~\ref{EQN:GPC_8} and \ref{EQN:GPC_30}) are violated as well. From an H-H distance of 1.294 \AA~-- 1.296 \AA, one occupation number is slightly negative ($\approx -10^{-3})$, but at a distance of 1.297 \AA, the 1RDM is ensemble $N$-representable again. Despite this, the state is clearly not pure-state $N$-representable because a staggering 14 GPCs are violated (Eqs.~\ref{EQN:GPC_41}--\ref{EQN:GPC_30} plus those listed in Table \ref{TAB:14_GPCS}).
\begin{table}[!htpb]
\caption{$\wedge^3 \mathcal{H}_{10}$ GPCs violated for H$_5^{2+}$ arranged in a regular pentagon, with an H-H distance of 1.297 \AA. Note that Eqs.~\ref{EQN:GPC_41}--\ref{EQN:GPC_30} are also violated at this geometry.}
\label{TAB:14_GPCS}  
    \centering
    {\footnotesize 
\begin{tabular}{rcl}
        \hline\hline
  $7 {\color{black}n}_1   +2 {\color{black}n}_2  -3 {\color{black}n}_3  -3 {\color{black}n}_4  +2 {\color{black}n}_5  +7 {\color{black}n}_6  -8{\color{black}n}_7  -3{\color{black}n}_8  -3{\color{black}n}_9  +2{\color{black}n}_{10}$ & $\le$ &6 \\ 
  $7 {\color{black}n}_1   +7 {\color{black}n}_2  -8 {\color{black}n}_3  +2 {\color{black}n}_4  -3 {\color{black}n}_5  -3 {\color{black}n}_6  +2{\color{black}n}_7  -3{\color{black}n}_8  -3{\color{black}n}_9  +2{\color{black}n}_{10}$ & $\le$ &6 \\ 
  $7 {\color{black}n}_1   +7 {\color{black}n}_2  -8 {\color{black}n}_3  +2 {\color{black}n}_4  -3 {\color{black}n}_5  -3 {\color{black}n}_6  -3{\color{black}n}_7  +2{\color{black}n}_8  +2{\color{black}n}_9  -3{\color{black}n}_{10}$ & $\le$ &6 \\ 
  $9 {\color{black}n}_1   -1 {\color{black}n}_2  -1 {\color{black}n}_3  -1 {\color{black}n}_4  -1 {\color{black}n}_5  +9 {\color{black}n}_6 -11{\color{black}n}_7  -1{\color{black}n}_8  -1{\color{black}n}_9  -1{\color{black}n}_{10}$ & $\le$ &7 \\ 
 $13 {\color{black}n}_1   +3 {\color{black}n}_2  -7 {\color{black}n}_3  +3 {\color{black}n}_4  -7 {\color{black}n}_5  +3 {\color{black}n}_6  -7{\color{black}n}_7  +3{\color{black}n}_8  -7{\color{black}n}_9  +3{\color{black}n}_{10}$ & $\le$ &9 \\ 
 $14 {\color{black}n}_1   +9 {\color{black}n}_2 -11 {\color{black}n}_3  -6 {\color{black}n}_4  +4 {\color{black}n}_5  +9 {\color{black}n}_6 -11{\color{black}n}_7  -6{\color{black}n}_8  -1{\color{black}n}_9  -1{\color{black}n}_{10}$ & $\le$ &12 \\ 
  $9 {\color{black}n}_1  +14 {\color{black}n}_2 -11 {\color{black}n}_3  -6 {\color{black}n}_4  +9 {\color{black}n}_5  +4 {\color{black}n}_6 -11{\color{black}n}_7  -6{\color{black}n}_8  -1{\color{black}n}_9  -1{\color{black}n}_{10}$ & $\le$ &12 \\ 
 $14 {\color{black}n}_1   +9 {\color{black}n}_2 -11 {\color{black}n}_3   9 {\color{black}n}_4 -11 {\color{black}n}_5  -6 {\color{black}n}_6  +4{\color{black}n}_7  -6{\color{black}n}_8  -1{\color{black}n}_9  -1{\color{black}n}_{10}$ & $\le$ &12 \\ 
 $23 {\color{black}n}_1  +13 {\color{black}n}_2 -17 {\color{black}n}_3  -7 {\color{black}n}_4  +3 {\color{black}n}_5 +13 {\color{black}n}_6 -17{\color{black}n}_7  -7{\color{black}n}_8  +3{\color{black}n}_9  -7{\color{black}n}_{10}$ & $\le$ &19 \\ 
 $13 {\color{black}n}_1  +23 {\color{black}n}_2 -17 {\color{black}n}_3  -7 {\color{black}n}_4 +13 {\color{black}n}_5  +3 {\color{black}n}_6 -17{\color{black}n}_7  -7{\color{black}n}_8  +3{\color{black}n}_9  -7{\color{black}n}_{10}$ & $\le$ &19 \\ 
 $23 {\color{black}n}_1  +13 {\color{black}n}_2 -17 {\color{black}n}_3 +13 {\color{black}n}_4 -17 {\color{black}n}_5  -7 {\color{black}n}_6  +3{\color{black}n}_7  -7{\color{black}n}_8  +3{\color{black}n}_9  -7{\color{black}n}_{10}$ & $\le$ &19 \\
        \hline\hline
\end{tabular}
}
\end{table}
Beyond 1.297 \AA, the 1RDM again loses ensemble $N$-representability, exhibiting, at various distances, occupation numbers that are less than zero (as negative as $-7.8$), greater than one (as large as 8.8), or complex-valued. Interestingly, the loss of ensemble-state $N$-representability in state 6 occurs near the ground-state equilibrium geometry (1.306 \AA), where one might expect that the ground-state to be well described by CCSD (especially since this is a three-electron system).

\subsection{$N$-representability in time-dependent EOM-CCSD}

\begin{figure}
    \centering
    \includegraphics{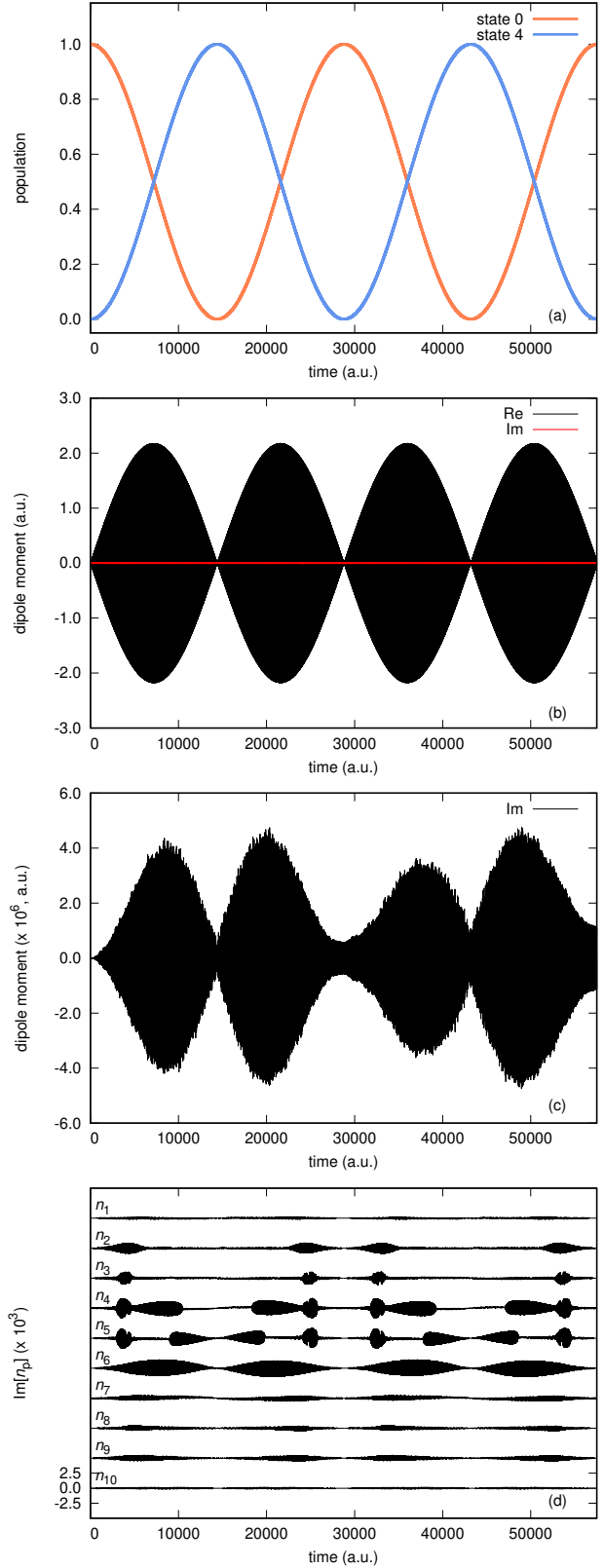}
    \caption{The (a) ground-state (state 0) and excited-state (state 4) populations, (b) dipole moment, (c) imaginary component of the dipole moment, and (d) imaginary component of the natural spin orbital occupation numbers, ${\color{black}n}_p$, for H$_5$ driven by an external field resonant with the state 0 $\to$ state 4 transition. }
    \label{FIG:H5_RABI}
\end{figure}

\begin{figure}
    \centering
    \includegraphics{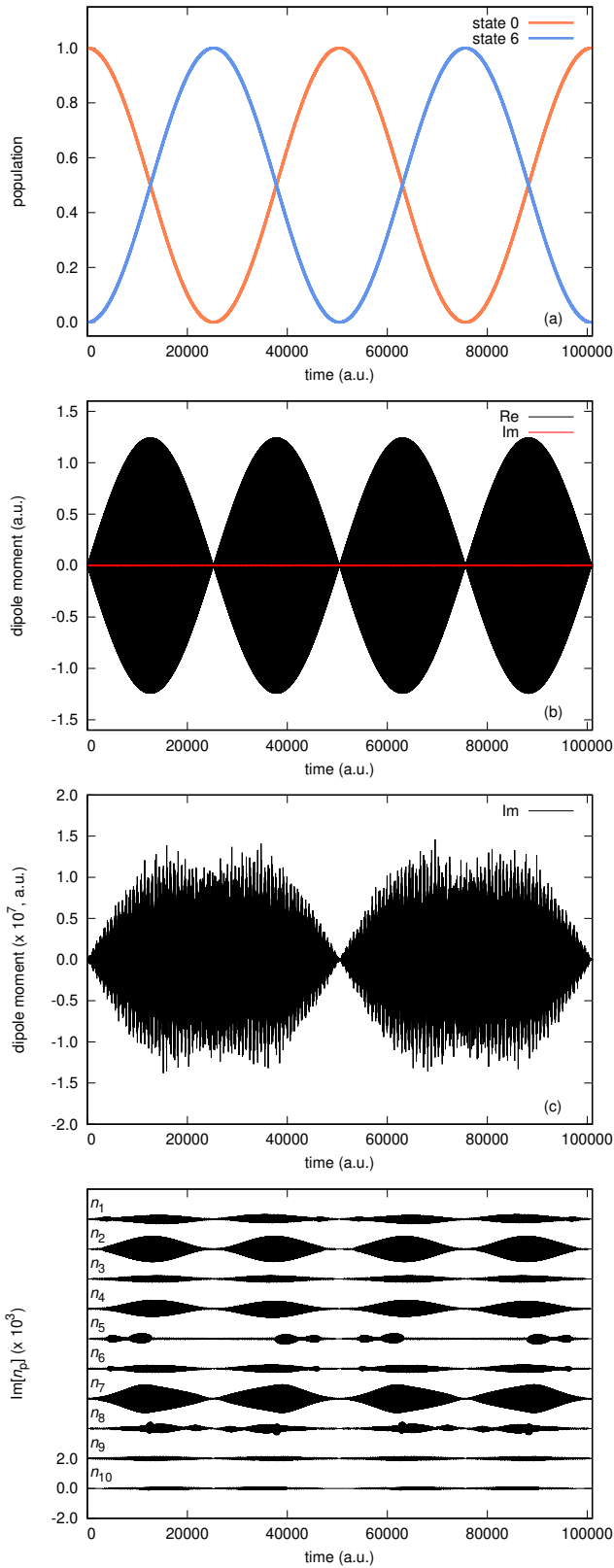}
    \caption{The (a) ground-state (state 0) and excited-state (state 6) populations, (b) dipole moment, (c) imaginary component of the dipole moment, and (d) imaginary component of the natural spin orbital occupation numbers, ${\color{black}n}_p$, for H$_5^{2+}$ driven by an external field resonant with the state 0 $\to$ state 6 transition. }
    \label{FIG:H5_2+_RABI}
\end{figure}

Lastly, we consider TD-EOM-CCSD calculations in which the system is driven between the ground state and an excited state by a resonant external electric field {\color{black} (\emph{i.e.}, Rabi flopping\cite{Merlin21_26}).} For these studies, we have chosen geometries at which both the ground state and the target excited state have pure-state $N$-representable 1RDMs, and the oscillator strength for the excited state is non-negative. So, by all appearances, the states in question seem physical. Figure \ref{FIG:H5_RABI}(a) depicts the populations of the ground state (state 0) and state 4 for the geometry depicted in Fig.~\ref{FIG:H5_LINEAR}, with $a$ = 0.45 \AA. The frequency of the external field is resonant with state 0 $\to$ state 4 transition; $\epsilon(t) = A {\rm sin}(\omega t) \hat{e}$, where {\color{black}$A = 10^{-4}$}, $\omega = 0.5135$ E$_{\rm h}$, and $\hat{e}$ is a unit vector polarized along the transition dipole moment for state 4. The populations in panel (a) show that the simulation covers roughly two Rabi cycles. Figure \ref{FIG:H5_RABI}(b) depicts the time-evolution of the dipole moment over the same period, which exhibits rapid oscillations with the external field and a beating characteristic of the Rabi cycle. The dipole moment becomes complex during the simulation, but the magnitude of the imaginary component of the dipole moment is small, never exceeding {\color{black}5$\times 10^{-6}$} a.u [panel (c)]. 
{\color{black}The imaginary component of the dipole displays a beating that is similar to that of the real component, except that it is not symmetric over the first and second halves of the Rabi cycle or over the first and second cycles.}
Panel (d) depicts the imaginary component of the natural spin-orbital occupation numbers along the two Rabi cycles, where we can see that the occupation numbers {\color{black}almost immediately} become complex{\color{black}, with the maximum magnitude of the imaginary components on the order of 10$^{-3}$. Ensemble $N$-representability of the 1RDM is never exactly recovered once it is lost, but there are some points in time when the imaginary components of all occupation numbers are smaller in magnitude than 10$^{-5}$. These points coincide with the system having been driven nearly entirely to either the ground or excited state. }

Figure \ref{FIG:H5_2+_RABI} depicts data for a similar time-dependent study on  H$_5^{2+}$ in which the system is driven from the ground state to state 6 by a resonant external electric field. Again, we have chosen a geometry at which the 1RDM for state 6 is pure-state $N$-representable: the regular pentagon considered above, with an H-H separation of 1.0 \AA.  The frequency of the external field is resonant with state 6; $\epsilon(t) = A {\rm sin}(\omega t) \hat{e}$, where {\color{black}$A = 10^{-4}$}, $\omega = 0.5528$ a.u., and $\hat{e}$ is a unit vector polarized along the transition dipole moment for state 6. Figure \ref{FIG:H5_2+_RABI}(a) depicts the populations of the ground state and state 6 as a function of time, which, again, covers roughly two Rabi cycles. 
Figure \ref{FIG:H5_2+_RABI}(b) depicts the time-evolution of the dipole moment over the same period, which exhibits the same beating seen in Fig.~\ref{FIG:H5_RABI}. The dipole moment becomes complex during the cycle, but the magnitude of the imaginary component is much smaller than in the case discussed above; it never exceeds {\color{black}1.5$\times 10^{-7}$} a.u. [panel (c)]. In this case, 
the beating in the imaginary part of the dipole moment is maximal when the system is in the excited state, which is different than the case in Fig.~\ref{FIG:H5_RABI}(c).
As was observed in Fig.~\ref{FIG:H5_RABI}(d), Fig.~\ref{FIG:H5_2+_RABI}(d) shows that the natural orbital occupation numbers  become complex valued {\color{black}almost immediately. As above, ensemble $N$-representability is never fully recovered, but the magnitude of imaginary components of the occupation numbers dip below 10$^{-5}$ in regions where population transfer is near complete.  }

\section{Conclusions}
\label{SEC:CONCLUSIONS}

While formulated in terms of right- and left-hand wave functions, truncated CC and EOM-CC  {\color{black}methods differ from Hermitian wave function theories in that} there is no guarantee that the RDMs derived from the left- and right-hand eigenfunctions of a similarity-transformed Hamiltonian expanded in an incomplete many-electron basis are derivable from any {\color{black}physical $N$-electron state}.
Indeed, we have demonstrated numerically that this is the case via analysis of the 1RDM, where severe violations in both ensemble-state\cite{Coleman63_668} and pure-state\cite{Klyachko08_287} $N$-representability conditions have been observed. Non-$N$-representable RDMs are common in the vicinity of avoided crossings, with particularly severe issues emerging near regions where the energy becomes complex.  We have also identified many states that exhibit negative oscillator strengths, which are clearly unphysical, at geometries where the 1RDMs for the ground and excited states are pure-state $N$-representable. In these cases, either the ground state or the excited state cannot be $N$-representable, so it is likely that additional ensemble-state\cite{Percus64_1756,Erdahl78_697,Percus04_2095, Mazziotti12_263002, Mazziotti12_062507, Mazziotti23_153001} or pure-state\cite{Mazziotti16_032516, DePrince22_5966} conditions on higher RDMs ({\em e.g.}, the two-electron RDM) associated with these states are violated.

We have also shown that unphysical 1RDMs arise in time-dependent simulations wherein an external field drives the system between two states with pure-state $N$-representable 1RDMs. This particular observation is interesting because it is generally assumed that wave function methods such as TD-EOM-CC are more reliable than less computationally demanding approaches such as real-time time-dependent density functional theory for modeling driven electron dynamics. Yet, we have identified situations where the 1RDM associated with a driven TD-EOM-CC state is not physical, even when the state is a linear combination of two EOM-CC states with pure-state $N$-representable 1RDMs. These observations call into question the reliability of properties evaluated using TD-EOM-CC-derived RDMs. Future studies should consider whether similar issues arise in time-dependent CC simulations that do not rely on the EOM-CC framework.\cite{Kvaal12_194109, Koch20_023115, Ishikawa18_051101, Pedersen20_071102,Crawford22_5479, Koch23_2301.05546,Pedersen23_e1666}

{\color{black}Lastly, we close by briefly considering how the nature of the reference configuration can impact the emergence of the unphysical features we have reported in this work. One may wonder, for example, whether these problems persist with the use of a spin-symmetry-broken (unrestricted Hartree-Fock, UHF) reference. To address this question, we performed additional tests using unrestricted EOM-CCSD calculations for the H$_5$/STO-3G system, along the dissociation path depicted in Fig.~\ref{FIG:H5_LINEAR}. The use of a UHF reference does mitigate the issues we have observed, but only in part. Many states with non-ensemble-state $N$-representable 1RDMs that have occupation numbers that are larger than one, less than zero, or complex can be found, but we do not observe any states with ensemble-state $N$-representable 1RDMs that violate the GPCs. We also do not observe any complex energy eigenvalues. We stress, however, that spin-symmetry breaking is not a panacea. Aside from the obvious loss of an important symmetry, there exist other systems for which the nature of the reference configuration does not impact the emergence of complex energy eigenvalues in truncated EOM-CC calculations (see Ref.~\citenum{DePrince23_2305.06412}, for example).}


{\color{black}
\section*{Appendix: EOM-CC as an RDM theory}

Here, we cast EOM-CC theory as an RDM theory using a constrained search formalism.\cite{Levy79_6062,Levy82_1200} We begin by considering the variance of the energy for the $I$-th state
\begin{equation}
    \sigma_I^2 = \langle \Psi_I | \hat{H}^2 | \Psi_I \rangle - \langle \Psi_I | \hat{H} | \Psi_I \rangle ^2
\end{equation}
and introducing a linear parametrization for the bra and ket states
\begin{align}
|\Psi_I \rangle &= \hat{C}_I | \Phi_0 \rangle \\
\langle \Psi_I | &= [\hat{C}_I | \Phi_0 \rangle ]^\dagger 
\end{align}
with 
\begin{equation}
\hat{C}_I = c_0 + \sum_{ia} c_i^a \hat{a}^\dagger_a \hat{a}_i + \frac{1}{4} \sum_{ijab} c_{ij}^{ab} \hat{a}^\dagger_a \hat{a}^\dagger_b \hat{a}_j \hat{a}_i + ... 
\end{equation}
We now define a functional of the 1RDM that minimizes $\sigma_I^2$ for all $|\Psi_I\rangle$ that map onto a fixed ${}^1{\bf D}$ ({\em i.e.}, those for which $\langle \Psi_I | \hat{a}^\dagger_p \hat{a}_q | \Psi_I \rangle = {}^1D^p_q$)
\begin{equation}
\label{EQN:1RDM_FUNCTIONAL}
F[{}^1{\bf D}] = \min_{|\Psi_I\rangle\to {}^1{\bf D}} ~ \sigma_I^2
\end{equation}
A search over all properly normalized, $N$-representable 1RDMs that minimize $F[{}^1{\bf D}]$ will yield a 1RDM (and normalized bra and ket states) for an approximate eigenfunction of $\hat{H}$; if $F[{}^1{\bf D}] = 0$, then the 1RDM corresponds to an eigenfunction of $\hat{H}$.
1RDMs (and wave functions) for additional states could be determined by also enforcing orthogonality of the parametrized states, {\em i.e.}, $\langle \Psi_I | \Psi_J \rangle = 0,~\forall I\neq J$. 
If the excitation order of $\hat{C}_I$ is not truncated below $N$, this constrained search formalism will be equivalent to solving the full CI problem, which is intractable in general. We could arrive at a more computationally feasible procedure by truncating the excitation order of $\hat{C}_I$, but the energies associated with the resulting truncated CI states would not have the correct scaling properties with system size.

A different parametrization of the 1RDM could be obtained by lifting the requirement that $\langle \Psi_I| = |\Psi_I\rangle ^\dagger$ and instead considering the biorthogonal variance expression
\begin{equation}
    \tilde{\sigma}_I^2 = \langle \tilde{\Psi}_I | \hat{H}^2 | \Psi_I \rangle - \langle \tilde{\Psi}_I | \hat{H} | \Psi_I \rangle ^2
\end{equation}
with $\langle \tilde{\Psi}_I|$ and $|\Psi_I\rangle$ defined by Eqs.~\ref{EQN:EOM_CC_RIGHT_WFN} and \ref{EQN:EOM_CC_LEFT_WFN}, respectively. We define a new functional of the 1RDM that minimizes $\tilde{\sigma}_I^2$ over all $\hat{L}_I$ and $\hat{R}_I$ that map onto ${}^1{\bf D}$ (given fixed $\hat{T}$ optimized for ground-state CC theory), {\em i.e.},
\begin{equation}
\label{EQN:1RDM_FUNCTIONAL_EOM_CC}
\tilde{F}[{}^1{\bf D}] = \min_{\hat{L}_I, \hat{R}_I \to {}^1{\bf D}} ~ \tilde{\sigma}_I^2
\end{equation}
Now, we can find ${}^1{\bf D}$ (and $\langle\tilde{\Psi}_I|$ and $|\Psi_I\rangle$) for different states by minimizing $\tilde{F}[{}^1{\bf D}]$
over the space of properly normalized, $N$-representable 1RDMs, while also enforcing the biorthogonality of the parametrized states, {\em i.e.}, $\langle \tilde{\Psi}_I | \Psi_J \rangle = 0,~\forall I\neq J$.  Without truncating $\hat{L}_I$ and $\hat{R}_I$, this procedure will be equivalent to full EOM-CC or the full CI and will be intractable. A more efficient protocol can be defined by truncating $\hat{L}_I$ and $\hat{R}_I$ at an excitation order less than $N$, in which case, unlike the truncated CI expansion, the energies associated with the EOM-CC states should have the correct scaling properties. 
Here, we note that truncated $\hat{L}_I$ and $\hat{R}_I$ produce non-symmetric 1RDMs, so the domain of ${}^1{\bf D}$ in the constrained search should be limited to those non-symmetric 1RDMs whose eigenvalues satisfy the $N$-representability conditions. Now, solutions to the truncated EOM-CC eigenvalue problem (Eqs.~\ref{EQN:EOM_CC_eig_R} and \ref{EQN:EOM_CC_eig_L}) should minimize $\tilde{\sigma}_I^2$ (giving $\tilde{\sigma}_I^2 = 0$), but the 1RDMs to which $\hat{R}_I$ and $\hat{L}_I$ correspond are not guaranteed to satisfy relevant $N$-representability conditions. In other words, the domain of 1RDMs obtained from solving the truncated EOM-CC eigenvalue problem is larger than the domain of those that can be obtained via the constrained search. Hence, unlike in the CI case, the constrained search is not strictly equivalent to solving the truncated EOM-CC equations directly for $\hat{R}_I$ and $\hat{L}_I$. To make the constrained search formalism equivalent to truncated EOM-CC eigenvalue problem, we must lift the requirement that ${}^1{\bf D}$ be $N$-representable when minimizing $\tilde{F}[{}^1{\bf D}]$. As such, through this lens, truncated EOM-CC theory can be viewed as an incomplete RDM theory that fails to guarantee that the biorthogonal parametrization of $\langle \tilde{\Psi}_I|$ and $|\Psi_I\rangle$ produces RDMs that are $N$-representable.  

Note that we could have also cast this problem as an optimization over higher-order RDMs ({\em e.g.}, the two-electron RDM, ${}^2{\bf D}$) that derive from $\hat{L}_I$ and $\hat{R}_I$. In this case, it would become clear that the truncated EOM-CC eigenvalue equations do not guarantee the $N$-representability of these higher-order RDMs either.}

\vspace{0.5cm}

{\bf Supplementary Material} Non-hermiticity of EOM-CCSD 1RDMs for the H$_5$/STO-3G system.

\vspace{0.5cm}

\begin{acknowledgments}This material is based upon work supported by the U.S. Department of Energy, Office of Science, Office of Advanced Scientific Computing Research and Office of Basic Energy Sciences, Scientific Discovery through the Advanced Computing (SciDAC) program under Award No. DE-SC0022263.\\ 
\end{acknowledgments}

\noindent {\bf DATA AVAILABILITY}\\

The data that support the findings of this study are available from the corresponding author upon reasonable request.

\bibliography{bib/Journal_Short_Name,bib/cc,bib/main,bib/deprince,bib/rdm}

\end{document}